# "Extra Dimensions and Compositeness as a Basis for Hierarchy in Quark Mass Matrices"

Bipin R. Desai  and  Alexander R. Vaucher

Department of Physics, University of California
Riverside, California 92521, USA
(September, 2000)

## Abstract

Strong gauge and top-Yukawa couplings predicted in the presence of extra dimensions lead to a trickle-down effect of the top-coupling through the renormalization group equations for the quark Yukawa matrices.  The matrix elements for the $u$ and $d$-quarks get progressively smaller as one moves away from the dominant (33)-element revealing a hierarchical pattern for the mass matrices.

The possible existence of extra dimensions has raised an intriguing question whether the gauge coupling constants $\alpha_i (= g_i^2/4\pi)$ become very large at some scale [1]. If, indeed, that happens then it makes sense that the quark-Yukawa couplings, particularly the largest of them, the top-coupling should behave the same way [1], [2].

A simple way to motivate this and to obtain $g_t$ in the process is to first consider the behavior of $\alpha_3$, the largest of the three gauge couplings, and relate it to the top coupling. Beyond $\mu = \mu_0$, ($\mu_0 = R^{-1}$, $R$ being the radius of the extra dimension), $\alpha_3$ is given by a power law [1]

$$\alpha_3^{-1}(\mu) = \alpha_3^{-1}(M_z) + \frac{3}{2\pi}\ln\left(\frac{\mu}{M_z}\right) + \frac{\widetilde{b}_3}{2\pi}\ln\left(\frac{\mu}{\mu_0}\right) - \frac{\widetilde{b}_3 X_\delta}{2\pi\delta}\left[\left(\frac{\mu}{\mu_0}\right)^\delta - 1\right] \qquad (1)$$

where $\delta$ represents additional dimensions, $X_\delta$ is a known function of $\delta$, and $\widetilde{b}_3 = -6 + 4\eta$, where $\eta$ represents the generation of chiral fermions of the MSSM. For $\eta \geq 2$, $\alpha_3^{-1}(\mu)$ decreases rapidly for $\mu > \mu_0$. Specifically, for $\eta = 2$, and $\delta = 1$ ($X_\delta = 2$) one can write (1) in this region as

$$\alpha_3^{-1} = c_3 - \frac{2}{\pi}e^t \qquad (2)$$

where $c_3$ is a known quantity and $t = \ln(\mu/\mu_0)$. Expanding near the zero at $t = \bar{t}_0$, we can write

$$\alpha_3^{-1} = \frac{2}{\pi}e^{\bar{t}_0}(\bar{t}_0 - t) \qquad (3)$$

If $g_t$ is the top-Yukawa coupling, and $\alpha_t = g_t^2/4\pi$, then in MSSM

$$\frac{d}{dt}\alpha_t^{-1} = -\frac{3}{\pi} + \frac{8}{3\pi}\frac{\alpha_t^{-1}}{\alpha_3^{-1}} \qquad (4)$$

Clearly then, because of (3), the right hand side would develop a pole and become infinite at $t = \bar{t}_0$. This is untenable as $\alpha_t$ would have the rather unphysical behavior of diving down to zero at that point (that is $\alpha_t^{-1}$ becomes infinite). The sensible alternative is to have $\alpha_t^{-1}$ vanish at $t = \bar{t}_0$ as well (i.e. both couplings share the same singular point).

Writing

$$\alpha_t^{-1} = c_t(\bar{t}_0 - t) \tag{5}$$

we find from (3) and (4) that

$$\frac{d}{dt}\alpha_t^{-1} = -\frac{3}{\pi} + \frac{4c_t}{3}e^{-\bar{t}_0} \tag{6}$$

Because $\bar{t}_0$ is very large one can ignore the second term on the right so that

$$\frac{d}{dt}\alpha_t^{-1} = -\frac{3}{\pi} \tag{7}$$

therefore

$$\alpha_t^{-1} = \frac{3}{\pi}(\bar{t}_0 - t)$$

which implies $c_t = \frac{3}{\pi}$ in (5) justifying *a posteriori* neglecting the second right hand side term in (6). The top coupling is now given by

$$g_t = \frac{2\pi}{\sqrt{3(t_0 - t)}} \tag{8}$$

This result, not surprisingly, is identical to that of Bardeen et al for the compositeness condition obtained through the vanishing of the Higgs renormalization parameter, $Z_H$, at the composite scale ($Z_H \to 0; g_t \to \infty$) [4]. And it is independent of the manner in which the gauge couplings become large.

What we have said above simply re-iterates the predictions of strong top-dynamics in the presence of extra dimensions [1], [2], [5].

Since $\mu_0$ does not enter explicitly in expression (8) we re-define $t$ and $\bar{t}_0$ in what follows

$$t \to t = \ln(\mu/1\,Gev) \\ \bar{t}_0 \to t_0 = \ln(\Lambda/1\,Gev)$$ 

(9)

where following [4] we call, $\Lambda$, the composite scale.

Since we know the explicit form of $g_t$ near the boundary, and know that it is singular there, we can use Frobenius-type method to solve the renormalization group equations (RGE) for the quark-Yukawa couplings. Eventhough it has often been stated that the RGE equations can not shed important light on the hierarchy question [6], the knowledge of $g_t$ turns out to provide very interesting results [7].

We find a definite trickle-down effect of the top-coupling as one moves away from the (33)-element of the u-quark matrix ($U$) into the other matrix elements revealing a hierarchical pattern. The same thing happens to the d-quark matrix ($D$).

We will first consider the $U$-Matrix where, because of the dominance of the top-coupling, the (33) element, to an excellent approximation, is given by $g_t$ itself.

$$U = \begin{pmatrix} u_{11} & u_{12} & u_{13} \\ u_{21} & u_{22} & u_{23} \\ u_{31} & u_{32} & u_{33} \end{pmatrix}$$

(10)

We will choose $U$ to be symmetric, so that

$$u_{ij} = u_{ji}$$

and

$$u_{33} = g_t = \frac{2\pi}{\sqrt{3(t_0 - t)}}$$

(11)

The RGE for $U$, for MSSM as an example, is given by

$$16\pi^2 \frac{dU}{dt} = \left[ -\sum_i c_i g_i^2 + 3UU^+ + DD^+ + Tr\left(3UU^+\right) \right] U \tag{12}$$

In order to demonstrate the hierarchy, we write the RGE in terms of the ratios

$$r_{ij} = \frac{u_{ij}}{u_{33}} \qquad , \qquad \left( r_{ij} = r_{ji} \right) \tag{13}$$

This also has the effect of eliminating the contribution of the gauge couplings. If we define

$$x = 1 - \frac{t}{t_0} \tag{14}$$

then we obtain the following RGEs

$$\frac{dr_{11}}{dx} = \frac{1}{4x}\left(r_{11} - r_{13}^2\right) \tag{15a}$$

$$\frac{dr_{12}}{dx} = \frac{1}{4x}\left(r_{12} - r_{23}r_{13}\right) \tag{15b}$$

$$\frac{dr_{13}}{dx} = -\frac{1}{4x}r_{23}r_{12} \tag{15c}$$

$$\frac{dr_{22}}{dx} = \frac{1}{4x}\left(r_{22} - r_{23}^2\right) \tag{15d}$$

$$\frac{dr_{23}}{dx} = -\frac{1}{4x}\left(r_{13}r_{12} + r_{23}r_{22}\right) \tag{15e}$$

We expect $|r_{ij}| < 1$ (except $r_{33} = 1$). If we write

$$r_{ij} = c_{ij}x^{n_{ij}} \tag{16}$$

near $x$=0 then, since

$$u_{33} \sim x^{-\frac{1}{2}} \tag{17}$$

from (11), the exponents $n_{ij}$ will be positive, and because of the square-root singularity of $u_{33}$, they will be $0, \frac{1}{4}, \frac{1}{2}, \dots$ etc.

Inserting (16) in equations (15a) through (15e) we note that each equation will provide two relations. One will relate the $n_{ij}$'s through the equality of the powers of $x$ on both sides and the other will involve mostly the $c_{ij}$'s.

We find the following relations between the coefficients, keeping $c_{23}$ as the input, and keeping only the leading terms [8]

$$\left.\begin{array}{l} c_{22} = c_{23}^2 \\ c_{13} = -c_{12}c_{23} \\ c_{11} = -c_{12}^2 c_{23}^2 \end{array}\right\} \tag{18}$$

We note immediately that, since we expect

$$c_{ij} < 1,$$

there is a definite hierarchical pattern. This is even more evident when we examine the $n_{ij}$'s. The $U$-matrix reads, after we put in all the results, as (normalized to $u_{33} = 1$)

$$U = \begin{pmatrix} -c_{12}^2 c_{23}^2 x^{\frac{1}{2}} & c_{12}x^{\frac{1}{4}} & -c_{12}c_{23}x^{\frac{1}{4}} \\ c_{12}x^{\frac{1}{4}} & c_{23}^2 & c_{23} \\ -c_{12}c_{23}x^{\frac{1}{4}} & c_{23} & 1 \end{pmatrix} \tag{19}$$

near $x$=0.

We notice that as we move away from the (33)-element the matrix elements get progressively smaller until we reach the (11)-element which is the smallest. The quark masses can be estimated from the diagonal elements,

$$m_u : m_c : m_t \approx c_{12}^2 c_{23}^2 x^{\frac{1}{2}} : c_{23}^2 : 1 \tag{20}$$

which shows a definite hierarchical pattern.

One often expresses the mass hierarchy in terms of the CKM parameters. In order to connect to the CKM elements, therefore, we write the RGE for $V_{CKM}$ in MSSM which reads

$$\left.\begin{aligned}
16\pi^2 \frac{d}{dt}\ln(V_{12}) &= O(V_{12}^3) \\
16\pi^2 \frac{d}{dt}\ln(V_{ij}) &= -g_t^2 \,, \, (ij = 13,23,31,32)
\end{aligned}\right\} \tag{21}$$

where we keep only the most dominant term on the right.

We write the CKM matrix in the Wolfenstein representation

$$V_{CKM} = \begin{pmatrix} 1 - \frac{\lambda^2}{2} & \lambda & A\lambda^3(\rho - i\eta) \\ -\lambda & 1 - \frac{\lambda^2}{2} & A\lambda^2 \\ A\lambda^3(1 - \rho - i\eta) & -A\lambda^2 & 1 \end{pmatrix} \tag{22}$$

Because $V_{12}$ is very small, (21) implies that it remains a constant $\approx \lambda$. The constants $c_{ij}$ in (16) may, therefore, be related to $\lambda$,

$$c_{ij} = f_{ij}(\lambda) \tag{23}$$

Considering specifically the (23)-term in (22), to avoid complex elements involving $\rho$ and $\eta$, the relations (11) and (21) state that

$$\frac{dV_{23}}{dx} = -\frac{1}{12x}V_{23} \tag{24}$$

therefore,

$$V_{23} \sim x^{\frac{1}{12}} \tag{25}$$

From the Wolfenstein parameterization this implies, since $\lambda$ is a constant, that

$$A \sim x^{\frac{1}{12}} \tag{26}$$

The matrix (19) can be written as

$$U \sim \begin{pmatrix} -c_{12}^2 c_{23}^2 A^6 & c_{12} A^3 & -c_{12} c_{23} A^3 \\ c_{12} A^3 & c_{23}^2 & c_{23} \\ -c_{12} c_{13} A^3 & c_{23} & 1 \end{pmatrix} \tag{27}$$

which shows the hierarchical behavior without the $x$ dependence.

From (23) if we take $c_{12} \sim \lambda^2$, and $c_{23} \sim \lambda^2$ then the mass ratio (20) would be

$$m_u : m_c : m_t \;=\; A^6 \lambda^8 : \lambda^4 : 1$$

which is consistent with the standard hierarchical representation of the $U$-quark masses.

Turning now to the $D$-matrix

$$D = \begin{pmatrix} d_{11} & d_{12} & d_{13} \\ d_{21} & d_{22} & d_{23} \\ d_{31} & d_{32} & d_{33} \end{pmatrix} \tag{28}$$

the corresponding RGE is given by

$$16\pi^2 \frac{dD}{dt} = \left[ -\sum_i c_i' g_i^2 + 3DD^+ + UU^+ + Tr\left(3DD^+\right) \right] D \tag{29}$$

It can be shown from the above equation that

$$d_{ij} \neq d_{ji} \tag{30}$$

i.e. $D$ is not symmetric.

Once again, as in the case of $g_t$ and $u_{33}$, we conclude that, to an excellent approximation,

$$d_{33} = g_b \tag{31}$$

where $g_b$ is the b-quark coupling.

From equation (29), keeping the dominant $g_t$ contribution on the right it is easy to verify that

$$d_{33} \sim x^{-\frac{1}{12}} \tag{32}$$

which has a singularity at $x=0$, like $u_{33}$, but a much milder one.

If we write

$$\rho_{ij} = \frac{d_{ij}}{d_{33}} \tag{33}$$

then from (29) we obtain the following eight equations, where on the right we have kept only the most dominant terms

$$\frac{d\rho_{11}}{dx} = \frac{1}{12x}\left(\rho_{11} - r_{13}\rho_{31}\right) \tag{34a}$$

$$\frac{d\rho_{12}}{dx} = \frac{1}{12x}\left(\rho_{12} - r_{13}\rho_{32}\right) \tag{34b}$$

$$\frac{d\rho_{13}}{dx} = \frac{1}{12x}\left(\rho_{13} - r_{13}\right) \tag{34c}$$

$$\frac{d\rho_{21}}{dx} = \frac{1}{12x}\left(\rho_{21} - r_{23}\rho_{31}\right) \tag{34d}$$

$$\frac{d\rho_{22}}{dx} = \frac{1}{12x}\left(\rho_{22} - r_{23}\rho_{32}\right) \tag{34e}$$

$$\frac{d\rho_{23}}{dx} = \frac{1}{12x}\left(\rho_{23} - r_{23}\right) \tag{34f}$$

$$\frac{d\rho_{31}}{dx} = \frac{1}{12x}\left(\rho_{31}r_{23}\rho_{23} - r_{23}\rho_{21}\right) \tag{34g}$$

$$\frac{d\rho_{32}}{dx} = \frac{1}{12x}\left(\rho_{32}r_{23}\rho_{23} - r_{23}\rho_{22}\right) \tag{34h}$$

Once again expressing

$$\rho_{ij} = \bar{c}_{ij} x^{\bar{n}_{ij}} \tag{35}$$

we obtain from the earlier results for $r_{ij}$ (normalizing $d_{33} = 1$)

$$D = \begin{pmatrix} -\dfrac{1}{3} c_{12} c_{23} \bar{c}_{21} x^{\frac{1}{3}} & \dfrac{1}{3} c_{12} c_{23}^2 x^{\frac{1}{3}} & \dfrac{1}{2} c_{12} c_{23} x^{\frac{1}{4}} \\ \bar{c}_{21} x^{\frac{1}{12}} & \bar{c}_{22} x^{\frac{1}{12}} & c_{23} \\ -c_{23} \bar{c}_{21} x^{\frac{1}{12}} & -c_{23} \bar{c}_{22} x^{\frac{1}{12}} & 1 \end{pmatrix} \tag{36}$$

and

$$D \sim \begin{pmatrix} -\dfrac{1}{3} c_{12} c_{23} \bar{c}_{21} A^4 & \dfrac{1}{3} c_{12} c_{23}^2 A^4 & \dfrac{1}{2} c_{12} c_{23} A^3 \\ \bar{c}_{21} A & \bar{c}_{22} A & c_{23} \\ -c_{23} \bar{c}_{21} A & -c_{23} \bar{c}_{22} A & 1 \end{pmatrix} \tag{37}$$

Again, a definite hierarchical pattern is exhibited in the above solution because the matrix elements get progressively smaller through the contributions of $c_{ij}$, $\bar{c}_{ij}$ and the powers of $x$ or $A$. Furthermore, unlike the $U$-matrix, the dependence on $x$ or $A$ is much milder in $D$.

The above results are obtained near the composite scale $x=0$. However, because $c_{ij}$ and $\bar{c}_{ij}$'s are expected to be constants, matrices (27) and (37) containing the Wolfenstein parameter $A$ maybe continuable to lower energies. This remains to be explored.

One could try to get more quantitative in discussing our results e.g. by assuming certain specific dependence for the constants $c_{ij}$ and $\bar{c}_{ij}$ so that our matrices reproduce the representations of $U$ and $D$ that are normally given in terms of $\lambda$. And one could invoke texture zeroes and introduce special models. We will take these up in a later work.

What is quite remarkable, however, is that once we know the specific singular form of $g_t$ (and $g_b$) at the composite scale, an inherent hierarchical form for the Yukawa matrices emerges simply based on the RGEs.

We thank Dr. Utpal Sarkar for many helpful discussions.

This work was supported in part by the U.S. Department of Energy under contract No: DE-FG03-94ER40837.